\documentclass[twocolumn,showpacs,preprintnumbers,amsmath,amssymb]{revtex4}
\usepackage{graphicx}
\usepackage{dcolumn}
\usepackage{bm}
\begin{document}
\title{Einstein spaces in warped geometries in five dimensions}
\author{M. Ar\i k}
\email{arikm@boun.edu.tr}
 \author{A. Baykal}
 \email{baykala@boun.edu.tr}
\author{ M. C. \c{C}al\i k}
 \author{D. \c{C}iftci}
 \email{dciftci@boun.edu.tr}
 \author{\"O. Delice}
 \email{odelice@boun.edu.tr}
\affiliation{Department of Physics, Bo\~{g}azi\c{c}i University, 34342 Bebek, Istanbul, Turkey
}
\begin{abstract}
We investigate five dimensional Einstein spaces in warped
geometries
 from the point of view of the four dimensional physically
relevant Robertson-Walker-Friedman cosmological metric and the
Schwarzschild metric. We show that a four-dimensional cosmology
with a closed spacelike section and a cosmological constant can be
imbedded into five-dimensional flat space-time.
\end{abstract}
\maketitle
The general theory of relativity is an experimentally
well-tested theory. Among these tests, the Schwarzschild solution
has played a central role. For the cosmological solutions,
however, the situation is beginning to clarify with the
accumulation of relevant astrophysical data. On the one hand, a
simple, consistent, logical cosmology requires a spatially
maximally symmetric Robertson-Walker-Friedman cosmology with
closed spacelike sections ($k=1$). Recent
 observational evidence shows that we live in an expanding
closed universe with positive cosmological constant
\cite{CCgroups}. The maximally symmetric Einstein de-Sitter
solutions are good prototypes of such space-times since they
include the cosmological constant. However, the existence of the
cosmological constant is one of the deep mysteries in cosmology.

Since the Kaluza-Klein idea \cite{kaluzaklein}, there have been
many theories suggesting that the universe may have more than four
dimensions. Nowadays, the idea that our universe may be a
three-brane embedded in five dimensional universe is very popular
\cite{dvali}-\cite{Randall2}. For a recent review see \cite{Reviews}.

 The recent interest in the
Randall-Sundrum \cite{Randall}-\cite{Randall2}  and related
scenarios has brought into consideration warped geometries such
that four dimensional spacetime metric is multiplied by a warp
factor which only depends on the coordinate of the extra
dimension, namely
\begin{equation}\label{rsmetric}
ds^2_{(5)}=dw\otimes dw+b^2(w)\ {\eta_{\mu\nu}dx^\mu\otimes dx^\nu}
\end{equation}
where $b(w)=e^{-k|w|}$ is the warp factor, 
 k is a constant and $\eta_{\mu\nu}=diag(-1,1,1,1)$. In their
second scenario \cite{Randall2}, where the range of the extra
dimension $w$ is $-\infty <w<+\infty$,  we live on a four
dimensional infinitely thin shell (three brane). Notice that the
five dimensional Einstein tensor outside the brane satisfies the
Einstein equation with a cosmological constant:
\begin{equation}\label{EECC5}
^{(5)}G_{MN}+g_{MN} \Lambda_5 =0 , \ M,N=0,1,2,3,5
\end{equation}
and on $w=$constant hypersurfaces  4 dimensional Einstein tensor
of this metric satisfies
\begin{equation}\label{EECC4}
^{(4)}G_{\mu\nu}+g_{\mu\nu} \Lambda_4 =0 , \ \mu,\nu=0,1,2,3.
\end{equation}
where $\Lambda_5 =-6k^2$ and $\Lambda_4=0$. The full Einstein
tensor of the 5 dimensional space-time of the metric
(\ref{rsmetric}) is given by
\begin{equation}\label{branebulk}
^{(5)}G_{MN}=-\eta_{MN} \Lambda_5-6k \delta^\mu_M \delta^\nu_N
\eta_{\mu\nu} \delta(w).
\end{equation}

Motivated by these considerations, in this work we will calculate
five dimensional Einstein equations of the metric (\ref{rsmetric})
for arbitrary $b(w)$ in terms of the four dimensional quantities
originating from the four dimensional metric
\begin{equation}\label{4 metric}
ds^2_{(4)}=g_{\mu\nu}dx^\mu \otimes dx^\nu,
\end{equation}
 and $b(w)$. As in the Randall-Sundrum scenario we do not wish any
 matter sources to survive on 5 dimensional space-time except a
 possible five  dimensional cosmological constant.
Our most important conclusion will be that a four-dimensional
cosmological constant can be induced even
when the five-dimensional cosmological constant is zero. We require that only gravity
 can propagate in extra dimensions. Thus the five
 dimensional space is an Einstein space where the original
 Randall-Sundrum metric  will be one of the cases of our
 solutions. Then, as in the Randall-Sundrum scenario we impose
 reflection ($Z_2$) symmetry on the extra dimension $w$. This symmetry
 will make the derivatives of the metric discontinuous with
 respect to $w$ at the point of symmetry and we know from the thin
 shell formalism of General Relativity \cite{israelshell} that this discontinuity
 will give rise to a surface layer (thin shell - brane). The resulting
 five dimensional Einstein tensor will be of the form
 (\ref{branebulk}). Since in our solutions four dimensional part of
 the metric is same for every $w$, the brane tension (the term proportional to
 $\delta(w)$) is caused only by the jump of $b'(w)$ on the brane.

 After calculating the five dimensional metric in terms of the four dimensional
 metric, we first consider the four dimensional cosmological solutions of Einstein equations
 where the four dimensional   space-time is an Einstein space and the four dimensional hypersurface is devoid of
 matter except a four dimensional cosmological constant. We  tabulate all
 possible solutions we find in
Table \ref{table1}.

 We then consider the four dimensional metric to be given by spherically
symmetric static Schwarzschild solution. For this
metric we  also find all possible solutions $b(w)$ when five dimensional
metric is an Einstein space and collect them in Table \ref{table2}. Then we
will also make some comments on these solutions.

Our five dimensional metric ansatz can be written in an
orthonormal basis as:
\begin{eqnarray}
\label{ourmetric}
ds^2_{(5)}&=&
dw\otimes dw+b(w)^2 \times \Big\{g_{\mu\nu}(x^\rho)dx^\mu\otimes
dx^\nu\Big\}\\
&=&\eta_{AB}E^A\otimes E^B
\end{eqnarray}
where the four dimensional metric is also written in an
orthonormal basis:
\begin{equation}\label{4metric}
ds^2_{(4)}=g_{ij}dx^i\otimes dx^j=\eta_{ij}e^i\otimes e^j.
\end{equation}
 The orthonormal basis one forms are chosen as
\begin{eqnarray}
E^i=b(w)e^i,\quad E^4=ibe^4=ibdt,\quad  E^5=dw ,
\end{eqnarray}
Note that for the sake of computational simplicity, we chose the timelike one form imaginary
so that we can take $\eta_{AB}$ as $\delta_{AB}$ and $\eta_{ij}$ as $\delta_{ij}$.
The indices run as $A,B,\ldots=1,2,3,4,5$ and $i,j=1,2,3,4$.

Employing Cartan  structure equations, we find the nonzero
components of the Riemann tensor as follows:
\begin{equation}\label{r5mnlk}
^{(5)}R^{ij}_{\phantom{ij}kl}=\frac{^{(4)}R^{ij}_{\phantom{ij}kl}}{b^2}-
\delta^{ij}_{kl}\frac{b'^2}{b^2},\ ^{(4)}R^{i5}_{\phantom{tt}i
5}=-\frac{b''}{b},
\end{equation}
where the $'$
on the functions denote derivatives of the functions with respect
to their arguments, and $\delta^{ij}_{kl}$ is generalized Kronecker delta.
 The Ricci  curvature scalar is found as:
\begin{eqnarray}\label{r5}
^{(5)}R=\frac{^{(4)}R}{b^2}-8\frac{b''}{b}-12\frac{b'^2}{b^2}, 
\end{eqnarray}
Using these one can easily calculate the nonzero components of the
Einstein tensor $G^{(5)}_{AB}$ of the metric (\ref{ourmetric}) as:
\begin{eqnarray}
^{(5)}G_{ij}&=&\frac{^{(4)}G_{ij}}{b^2}+\delta_{ij}\left\{3\frac{b''}{b}+3\frac{b'^2}{b^2}\right\},\\
^{(5)}G_{55}&=&-\frac{R^{(4)}}{2b^2}+\frac{6b'^2}{b^2}.
\end{eqnarray}
We have calculated the nonzero components of the five dimensional
Einstein tensor for the metric (\ref{ourmetric}) in terms of
$b(w)$ and the four dimensional Einstein tensor of the metric
(\ref{4metric}). Since  we want to first investigate the
cosmological solutions we chose four dimensional metric ansatz as
follows:
\begin{equation}
\label{ds4}ds^2_{(4)}=-dt^2+a(t)^2 ds^2_{3},
\end{equation}
where
\begin{equation}\label{ds3}
ds^2_{(3)}=d\chi^2+c(\chi)^2 d\Omega^2_2,\ \
d\Omega^2_2=d\theta^2+sin^2\theta d\phi^2.
\end{equation}
Here we will find admissible values of $b(w),a(t),c(\chi)$ when the
Einstein equations satisfy (\ref{EECC5}) and (\ref{EECC4}). We can
read off the orthonormal basis one forms $e^i$ from (\ref{ds4}) and
(\ref{ds3}):
\[
\begin{array}{lc}
e^i=\{e^4,a(t)e^a\}, \qquad e^4=idt,&\\[.1cm]
 e^a=\{d\chi,\ c(\chi)d\theta,\;
c(\chi)sin\theta d\phi\},\qquad &a,b\ldots=1,2,3.
\end{array}
\]
For (\ref{r5mnlk}) the nonzero components of the four dimensional
Riemann tensor are found as:
\begin{equation}
^{(4)}R^{ab}_{\phantom{ab}cd}=\frac{^{(3)}R^{ab}_{\phantom{ab}cd}}{a^2}
+\delta^{ab}_{cd}\frac{\dot{a}^2}{a^2},
\qquad^{(4)}R^{a4}_{\phantom{qw}a4}=\frac{\ddot{a}}{a},
\end{equation}
and
\begin{equation}
^{(3)}R^{12}_{\phantom{12}12}=^{(3)} R^{13}_{\phantom{13}13}=-\frac{\check{\check{c}}}{c},
\qquad ^{(3)}R^{23}_{\phantom{23}23}=\frac{1-\check{c}^2}{c}.
\end{equation}
For the Ricci  curvature scalar (\ref{r5}), we have:
\begin{equation}
^{(4)}R=\frac{^{(3)}R}{a^2}+6\left\{\frac{\ddot{a}}{a}+\frac{\dot{a}^2}{a^2}
\right\}, \ \
^{(3)}R=-4\frac{\check{\check{c}}}{c}+2\frac{1-\check{c}^2}{c^2}
\end{equation}
The Einstein tensor for this metric (\ref{ds4}) is
\begin{eqnarray}
^{(4)}G_{ab}&=&\frac{^{(3)}G_{ab}}{a^2}-\delta_{ab}\left\{\frac{2\ddot{a}}{a}+\frac{\dot{a}^2}{a^2}\right\},\\
^{(4)}G_{44}&=&-\frac{R^{(3)}}{2a^2}-3\frac{\dot{a}^2}{a^2}
\end{eqnarray}
where
\begin{equation}
^{(3)}G_{11}=\frac{\check{c}^2-1}{c^2},\
^{(3)}G_{22}=^{(3)}G_{33}=\frac{\check{\check{c}}}{c}.
\end{equation}
Let us combine all these, then $^{(5)}G_{AB}$ becomes:
\begin{eqnarray}
\label{G11}^{(5)}G_{11}\!\!\!&=&\!\!\!\left\{ \frac{\check{c}^2-1}{a^2c^2}-
\left( 2\frac{\ddot{a}}{a}+\frac{\dot{a}^2}{a^2}\right) \right\}
\frac{1}{b^2}+\frac{3b''}{b}+\frac{3b'^2}{b^2}\phantom{AAAA} \\
\label{G22} ^{(5)}G_{22}&=&\left\{ \frac{\check{\check{c}}}{a^2c}-\left( 2\frac{\ddot{a}}{a}+\frac{\dot{a}^2}{a^2}\right) \right\}\frac{1}{b^2}+\frac{3b''}{b}+\frac{3b'^2}{b^2} \\
\notag&=&^{(5)}G_{33}\;  \\
\label{Gtt} ^{(5)}G_{44}&=&\left\{ \frac{2\check{\check{c}}}{a^2c}+ \frac{\check{c}^2-1}{a^2c^2}-3\frac{\dot{a}^2}{a^2} \right\}\frac{1}{b^2}+\frac{3b''}{b}+\frac{3b'^2}{b^2} \\
\label{G55}^{(5)}G_{55}&=&\left\{ \frac{2\check{\check{c}}}{a^2c}+
\frac{\check{c}^2-1}{a^2c^2}- 3(\frac{\ddot{a}}{a} +
\frac{\dot{a}^2}{a^2}) \right\}\frac{1}{b^2}+\frac{6b'^2}{b^2}.
\end{eqnarray}

As we said before, we want to solve these for $a$ and $b$ from
$^{(5)}G_{AB}+\delta_{AB}\Lambda_5=0$. For
 $^{(5)}G_{11}=^{(5)}G_{22}$ we get the following differential
equation:
\begin{equation}
\frac{\check{\check{c}}}{c}=\frac{\check{c}^2-1}{c^2}
\end{equation} whose set of solutions is
\begin{equation}
c(\chi)=\left\{\chi, \ \frac{1}{c_0}\sin(c_0\chi),\
\frac{1}{c_0}\sinh(c_0\chi) \right\},
\end{equation}
which correspond respectively to the cases $k=0,1,-1$.

For k=0, $^{(5)}G_{ii}= ^{(5)}G_{44}$ gives the following
differential equation:
\begin{equation}
\frac{\ddot{a}}{a}=\frac{\dot{a}^2}{a^2},
\end{equation}
whose set of solutions is
\begin{equation}
  a(t)=\left\{1; \; e^{a_0t}  \right\}.
\end{equation}
For the $a=1$ case, $^{(5)}G_{ii}=\ ^{(5)}G_{55}$ gives the
equation
\[
\frac{b''}{b}=\frac{b'^2}{b^2},
\]
whose set of solutions is
\[b(w)=\left\{1; \ e^{b_0w}
\right\}.
\]
Finally, for the  $a=e^{a_0t}$ case, we have
\[
\frac{b''}{b}=\frac{b'^2-a_0^2}{b^2},
\]
whose set of solutions is
\[
b(w)=\left\{a_0w;\ \frac{a_0}{b_0}\sinh(b_0w);\
\frac{a_0}{b_0}\sin(b_0w)    \right\}.
\]
In the same way, we can easily find the solutions
for $k=\mp1$ .
 All solutions are shown in  Table \ref{table1}. As in the Randall-Sundrum
 case,  to have a brane embedded in
 five dimensions for these solutions we have to impose $Z_2$ symmetry
 on $b(w)$.
 Then our four dimensional universe will be an infinitely thin shell
 at $w=0$ and
 the total five dimensional  Einstein tensor will have of the
 form:
 \begin{eqnarray}
 ^{(5)}_{(T)}G_{AB}&=&^{(5)}G_{AB}+6\frac{b'}{b}\delta(w)\notag\\
&=&-\Lambda_5\delta_{AB}-\sigma \delta_{AB}
\delta_{\mu}^{A}\delta_{\nu}^{B} \delta(w).
\end{eqnarray}
The $k=1$ case corresponds to  closed expanding universe with
positive cosmological constant, which is in accordance with recent
observations \cite{CCgroups}. For this case, Table (\ref{table1})
shows that $b(w)$ can take three different values: $\{w;\
\sinh w;\ \sin w\}$. The first of these is very interesting since
in this case the five-dimensional Riemann tensor and the
five-dimensional cosmological constant are zero.

To have a $k=1$ solution
with $^{(5)}R^{MN}_{\phantom{MN}PQ}=0$ for this geometry, it is necessary to
have nonzero $\Lambda_4$.
 So, flat and empty five-dimensional Minkowski universe in warped geometry
(\ref{ourmetric}) can give rise to a four dimensional closed
expanding universe with positive cosmological constant. Imposing
$Z_2$ symmetry, the metric for this case becomes:
\begin{eqnarray}
\label{metricwithrimtenzero}
& &ds^2_{(5)}=dw^2+
\left(a_0|w|\right)^2  \\
& \times & \left\{-dt^2+\frac{c_0^2}{a_0^2}
  \cosh^2(a_0t)
\Big\{d\chi^2+\frac{1}{c_0^2}\sin^2(c_0\chi)d\Omega_2^2
\Big\}\right\},\nonumber \
\end{eqnarray}
with
\begin{eqnarray}
^{(5)}G_{\mu\nu}&=&-6b'/b\ \delta_{\mu\nu} \delta(w), \nonumber \\
^{(5)}G_{55}&=&^{(5)}G_{5\mu}=0, \quad  ^{(4)}G_{\mu\nu}=-\Lambda_4
\delta_{\mu\nu}.
\end{eqnarray}
For this case the matter content of the  four dimensional universe
is only the four-dimensional cosmological constant.
In fact,
observations show that, the cosmological constant dominates the
matter content of the universe. According to the  recent review
\cite{padmanabhan}, the composition of the content of the universe is as follows:
%
\addtolength{\hoffset}{.5cm}
\begin{table*}
\begin{tabular}{|l|c|c|l|c|l|c|c|c|c|l|}
\hline
 &  & & &  &  &  &&  & & \\[-.3cm]
\multicolumn{1}{|c|}{$k$} & \multicolumn{1}{c|}{$c(\chi)$} & \multicolumn{1}{c|}{$a(t)$} & \multicolumn{1}{c|}{$b(w)$} & $R^{M(5)}_{NPQ}$ & $R^{(5)}$ & $\Lambda_5$ & $R^{\mu(4)}_{\nu\lambda\kappa}$ & $R^{(4)}$ & $\Lambda_4$ & $R^{(3)}$
\\[.1cm]
\hline
 &  & & &  &  &  &&  & & \\[-.3cm]
\multicolumn{1}{|c|}{} & \multicolumn{1}{c|}{} & \multicolumn{1}{c|}{} & \multicolumn{1}{c|}{1} & $0$ & \multicolumn{1}{c|}{0} & 0 & 0 & \multicolumn{1}{c|}{} & \multicolumn{1}{c|}{} & \multicolumn{1}{c|}{} \\[.1cm]
\cline{4-8}
&  &1 & &  &  &  && 0 &0 & \\[-.3cm]
\multicolumn{1}{|c|}{} & \multicolumn{1}{c|}{} & \multicolumn{1}{c|}{} & \multicolumn{1}{c|}{$e^{b_ow}$} & $-b_0^2$ & \multicolumn{1}{c|}{$-20b_0^2$} & $-6b_0^2$ & 0 & \multicolumn{1}{c|}{} & \multicolumn{1}{c|}{} & \multicolumn{1}{c|}{} \\[.1cm]
\cline{3-10}
&  & & &  &  &  &&  & & \\[-.3cm]
\multicolumn{1}{|c|}{0} & \multicolumn{1}{c|}{$\chi$} & \multicolumn{1}{c|}{} & \multicolumn{1}{c|}{$a_0w$} & $0$ & \multicolumn{1}{c|}{0} & 0 & $a_0^2$ & \multicolumn{1}{c|}{} & \multicolumn{1}{c|}{} & \multicolumn{1}{c|}{0} \\[.1cm]
\cline{4-8}
&  & & &  &  &  &&  & & \\[-.3cm]
\multicolumn{1}{|c|}{} & \multicolumn{1}{c|}{} & \multicolumn{1}{c|}{$e^{a_0t}$} & \multicolumn{1}{c|}{$\frac{a_0}{b_0}\sinh(b_0w)$} & $-b_0^2$ & \multicolumn{1}{c|}{$-20b_0^2$} & $-6b_0^2$ & $a_0^2$ & \multicolumn{1}{c|}{$12a_0^2$} & \multicolumn{1}{c|}{$3a_0^2$} & \multicolumn{1}{c|}{} \\[.1cm]
\cline{4-8}
&  & & &  &  &  &&  & & \\[-.3cm]
\multicolumn{1}{|c|}{} & \multicolumn{1}{c|}{} & \multicolumn{1}{c|}{} & \multicolumn{1}{c|}{$\frac{a_0}{b_0}\sin (b_0w)$} & $b_0^2$ & \multicolumn{1}{c|}{$20b_0^2$} & \multicolumn{1}{r|}{$6b_0^2$} & $a_0^2$ & \multicolumn{1}{r|}{} & \multicolumn{1}{r|}{} & \multicolumn{1}{r|}{} \\[.1cm]
\hline
&  & & &  &  &  &&  & & \\[-.3cm]
\multicolumn{1}{|c|}{} & \multicolumn{1}{c|}{} & \multicolumn{1}{c|}{} & \multicolumn{1}{c|}{$a_0w$} & $0$ & \multicolumn{1}{c|}{0} & 0 & $a_0^2$ & \multicolumn{1}{c|}{} & \multicolumn{1}{c|}{} & \multicolumn{1}{c|}{} \\[.1cm]
\cline{4-8}
&  & & &  &  &  &&  & & \\[-.3cm]
\multicolumn{1}{|c|}{1} & \multicolumn{1}{c|}{$\frac{1}{c_0}\sin(c_0\chi)$} & \multicolumn{1}{c|}{$\frac{c_0}{a_0}\cosh(a_0t)$} & \multicolumn{1}{c|}{$\frac{a_0}{b_0}\sinh(b_0w)$} & $-b_0^2$ & \multicolumn{1}{c|}{$-20b_0^2$} & $-6b_0^2$ & $a_0^2$ & \multicolumn{1}{c|}{$12a_0^2$} & \multicolumn{1}{c|}{$3a_0^2$} & \multicolumn{1}{c|}{$6c_0^2$} \\[.1cm]
\cline{4-8}
&  & & &  &  &  &&  & & \\[-.3cm]
\multicolumn{1}{|c|}{} & \multicolumn{1}{c|}{} & \multicolumn{1}{c|}{} & \multicolumn{1}{c|}{$\frac{a_0}{b_0}\sin(b_0w)$} & $b_0^2$ & \multicolumn{1}{c|}{$20b_0^2$} & $6b_0^2$ & $a_0^2$ & \multicolumn{1}{c|}{} & \multicolumn{1}{c|}{} & \multicolumn{1}{c|}{} \\[.1cm]
\hline
&  & & &  &  &  &&  & & \\[-.3cm]
\multicolumn{1}{|c|}{} & \multicolumn{1}{c|}{} & \multicolumn{1}{c|}{} & \multicolumn{1}{c|}{1} & $0$ & \multicolumn{1}{c|}{0} & 0 & 0 & \multicolumn{1}{c|}{} & \multicolumn{1}{c|}{} & \multicolumn{1}{c|}{} \\[.1cm]
\cline{4-8}
&  & $c_0t$& &  &  &  &&  0&0 & \\[-.3cm]
\multicolumn{1}{|c|}{} & \multicolumn{1}{c|}{} & \multicolumn{1}{c|}{} & \multicolumn{1}{c|}{$e^{b_0w}$} & $-b_0^2$ & \multicolumn{1}{c|}{$-20b_0^2$} & $-6b_0^2$ & 0 & \multicolumn{1}{c|}{} & \multicolumn{1}{c|}{} & \multicolumn{1}{c|}{} \\[.1cm]
\cline{3-10}
&  & & &  &  &  &&  & & \\[-.3cm]
\multicolumn{1}{|c|}{-1} & \multicolumn{1}{c|}{$\frac{1}{c_0}\sinh(c_0\chi)$} & \multicolumn{1}{c|}{} & \multicolumn{1}{c|}{$a_0w$} & $0$ & \multicolumn{1}{c|}{0} & 0 & $a_0^2$ & \multicolumn{1}{c|}{} & \multicolumn{1}{c|}{} & \multicolumn{1}{c|}{$-6c_0^2$} \\[.1cm]
\cline{4-8}
&  & & &  &  &  &&  & & \\[-.3cm]
\multicolumn{1}{|c|}{} & \multicolumn{1}{c|}{} & \multicolumn{1}{c|}{$\frac{c_0}{a_0}\sinh(a_0t)$} & \multicolumn{1}{c|}{$\frac{a_0}{b_0}\sinh(b_0 w)$} & $-b_0^2$ & \multicolumn{1}{c|}{$-20b_0^2$} & $-6b_0^2$ & $a_0^2$ & \multicolumn{1}{c|}{$12a_0^2$} & \multicolumn{1}{c|}{$3a_0^2$} & \multicolumn{1}{c|}{} \\
\cline{4-8}
&  & & &  &  &  &&  & & \\[-.3cm]
\multicolumn{1}{|c|}{} & \multicolumn{1}{c|}{} & \multicolumn{1}{c|}{} & \multicolumn{1}{c|}{$\frac{a_0}{b_0}\sin(b_0 w)$} & $b_0^2$ & \multicolumn{1}{c|}{$20b_0^2$} & $6b_0^2$ & $a_0^2$ & \multicolumn{1}{c|}{} & \multicolumn{1}{c|}{} & \multicolumn{1}{c|}{} \\[.1cm]
\cline{3-10}
 &  & & &  &  &  &&  & & \\[-.3cm]
\multicolumn{1}{|c|}{} & \multicolumn{1}{c|}{} & \multicolumn{1}{c|}{$\frac{c_0}{a_0}\sin (a_0t)$} & \multicolumn{1}{c|}{$\frac{a_0}{b_0}\cosh(b_0 w)$} & $-b_0^2$ & \multicolumn{1}{c|}{$-20b_0^2$} & $-6b_0^2$ & $-a_0^2$ & \multicolumn{1}{c|}{$-12a_0^2$} & \multicolumn{1}{c|}{$-3a_0^2$} & \multicolumn{1}{c|}{} \\[.1cm]
\hline
\end{tabular}
\caption{\scriptsize{$b(w),a(t),c(\chi)$ and other quantities for 5-d
 Einstein space when 4-d part is of the form}
(\ref{ourmetric})
 }\label{table1}
\end{table*}
\addtolength{\hoffset}{-.5cm}
\begin{eqnarray}
&&\Omega_B\approx(0.01-0.2), \quad \Omega_{R}\approx 2\times 10^{-5}, \nonumber \\
&& \Omega_{DM}\approx 0.3, \quad  \Omega_{\Lambda}\approx 0.7,
\end{eqnarray}
where $\Omega_B$ is the density parameter of the visible,
nonrelativistic, baryonic matter; $\Omega_R$ is the density
parameter of the radiation; $\Omega_{DM}$ is the density parameter
of the pressureless nonbaryonic dark matter
; and $\Omega_\Lambda$ is the density parameter of the
cosmological constant. According to the observations which use several
independent techniques, the density parameter of the
nonrelativistic matter is 
$\Omega_{NR}=(\Omega_{B}+\Omega_{(DM)})\approx(0.2 \sim 0.4)$.
  This raises the possibility whether with just a four dimensional cosmological constant
the five dimensional space-time is flat except on the brane. Other
kinds of
 matter in four dimensions require the five dimensional space-time to fluctuate from flat.
  Thus the presence of five dimensions differentiates between "dark energy"
satisfying equation of state $p=-\rho$ and other forms of
matter-energy. Although four dimensional De-Sitter space with
positive cosmological constant is consistent with five dimensional
flat space, other types of matter-energy in four dimensions
require the five dimensional space-time to fluctuate from flatness.

Now we turn to discuss the four dimensional
Schwarzschild solution from the five dimensional point of view.
Let us choose $ds^2_{(4)}$ as Schwarzschild de-Sitter metric which
satisfies (\ref{EECC4}) and
is given by:
\begin{eqnarray}
\label{schwdesitter}
ds^2_{(4)}&=&-\left\{1-\frac{2m}{r}-d_0 r^2
\right\}dt^2\nonumber \\
&+&\left\{1-\frac{2m}{r}-d_0 r^2\right\}^{-1}dr^2+r^2
d\Omega^2_2.
\end{eqnarray}
 We  find $b(w)$ for $d_0<0,d_0>0,d_0=0$ when
the metric (\ref{ourmetric}) satisfies (\ref{EECC5}) and presented
in the Table (\ref{table2}). Note that for $m=0$ and for $d_0>0$,
The Schwarzschild-de Sitter metric becomes a maximally symmetric
metric and this metric can be written in a form where spacelike
sections are closed. The metric (\ref{schwdesitter}) can be
transformed into:
\begin{table}[!hbp]
\begin{center}
\begin{tabular}{|c|c|c|c|c|}
\hline
 &  &  &  &  \\[-.3cm]
$\Lambda_4$ & $b(w)$ & $R^{(5)}$ & $\Lambda_5$ & $R^{(4)}$ \\[.1cm]
\hline
 &  &  &  &  \\[-.3cm]
$$ & $1$ & 0 & 0 &  \\[.1cm]
\cline{2-4}
 0&&  &  &  0\\[-.3cm]
 & $e^{b_0w}$ & $-20b_0^2$ & $-6b_0^2$ &  \\[.1cm]
\hline
 &  &  &  &  \\[-.3cm]
 &  $ w$ & 0 & $0$ &  \\
 &  &  &  &  \\[-.3cm]
\cline{2-4}
 &  &  &  &  \\[-.3cm]
$3d_0$ & $\frac{d_0}{b_0} \sinh (b_0w)$ & $-20b_0^2$ & $-6b_0^2$ & $12d_0^2$ \\
 &  &  &  &  \\[-.3cm]
\cline{2-4}
 &  &  &  &  \\[-.3cm]
 & $\frac{d_0}{b_0} \sin(b_0w)$ & $20b_0^2$ & $6b_0^2$ &  \\
 &  &  &  &  \\[-.3cm]
\hline
 &  &  &  &  \\[-.3cm]
$-3d_0$ & $\frac{d_0}{b_0} \cosh (b_0w)$ & $-20b_0^2$ & $-6b_0^2$
& $-12d_0^2$
\\[.04cm]
\hline
\end{tabular}
\caption{$b(w)$ \scriptsize{for different signs of $b_0$}}\label{table2}
\end{center}
\end{table}
\begin{equation}
ds^2_{(4)}=-dt'^2+\cosh^2(t')\left\{d\chi^2+ \sin^2\chi d\Omega_2^2
\right\}
\end{equation}
with the following transformation
\begin{eqnarray}
r&=&\cosh(t') \sin(\chi) ,\nonumber \\
t&=&\ln\left\{\frac{\sinh(t')+\cosh(t')\cos(\chi)}{\left\{1-\cosh^2(t')\sin^2(\chi)\right\}^{1/2}}
\right\} .
\end{eqnarray}

For this Schwarzschild-de Sitter case, for $b(w)\sim w$ and $m\neq
0$, five dimensional Riemann tensor is not zero or constant but
involves terms proportional to $m/r^3$. If $m=0$, the solution
reduces to (\ref{metricwithrimtenzero}). Thus, if we impose $Z_2$
symmetry, there will be a brane at $w=0$. Having matter sources on
the brane will change the five dimensional metric from flat to
curved. Five dimensional Ricci flat but curved metric in warped
geometry can give rise to a four dimensional universe with
positive cosmological constant and matter. This is a special case
of space-time matter (or induced matter) theorem
\cite{wessonstmbook} which states that the matter content of the
universe is induced from higher-dimensional geometry. The
relevance of this theorem has been emphasized from the R-S point
of view by Wesson and Seahra \cite{wessonseahra}.

In conclusion, we have shown that if in a Randall-Sundrum like
scenario one imposes the condition that $4+1$ dimensional
space-time is flat, the only $3+1$ dimensional brane which admits
a closed spacelike section cosmology requires a four dimensional
cosmological constant.
It is clear
from Table I that in fact all flat five-dimensional space-time
manifolds in warped geometries (\ref{ourmetric}) imply a nonzero and
positive cosmological constant for the four-dimensional cosmology.
This fact may be important as far as the measured
cosmological constant is positive.

\end{document}